\documentclass[showpacs,twocolumn,aps,prl]{revtex4}
\usepackage{amsmath}
\usepackage{ulem}
\usepackage{amsfonts}
\usepackage{amssymb}
\usepackage{amsthm}
\usepackage{graphicx}
\usepackage{CJK}
\usepackage{subfigure}
\usepackage{amsmath,bm}
\begin{document}

%----------------------------------------------------------------%
\title{On the instabilities of the Walker propagating domain wall solution}
\author{B. Hu}
\author{X.R. Wang}
\email{[Corresponding author:]phxwan@ust.hk}
\affiliation{Physics Department, The Hong Kong University of
Science and Technology, Clear Water Bay, Kowloon, Hong Kong}
\affiliation{HKUST Shenzhen Research Institute, Shenzhen 518057, China}
\date{\today}
\begin{abstract}
A powerful mathematical method for front instability analysis that was
recently developed in the field of nonlinear dynamics is applied to the
1+1 (spatial and time) dimensional Landau-Lifshitz-Gilbert (LLG) equation.
From the essential spectrum of the LLG equation,
it is shown that the famous Walker rigid body propagating domain
wall (DW) is not stable against the spin wave emission.
In the low field region only stern spin waves are emitted while
both stern and bow waves are generated under high fields.
By using the properties of the absolute spectrum of the LLG equation,
it is concluded that in a high enough field, but below the Walker
breakdown field, the Walker solution could be convective/absolute
unstable if the transverse magnetic anisotropy is larger than a
critical value, corresponding to a significant modification of the
DW profile and DW propagating speed. Since the Walker solution of
1+1 dimensional LLG equation can be realized in experiments, our
results could be also used to test the mathematical method in a
controlled manner.
\end{abstract}
% 05.45.-a Nonlinear dynamics and nonlinear dynamical systems
%reversal mechanisms, 75.60.Jk
%spin waves, 75.30.Ds
%magnetic properties and materials, 75.60.Ch
\maketitle

%-----------------------------------------------------------%
\section{I. INTRODUCTION}

The past century witnessed the quantum leap of the semiconductor industries
which gave birth to the computer science and information technology.
We are now in an era in which information keeps being generated at a
skyrocketing pace such that the net volume of information produced per
day might be comparable to that accumulated after years one century ago.
As an important participant, magnetic data recording now has assumed the
major task of information documentation, through video tapes, hard disks, etc.
In order to cope with the exponentially growing information volume,
the need to develop data storage devices with higher capacity and
faster read/write operation speed is demanded.
This intrigue the development of spintronics-the pursuit to employ, in
addition to the charge of electrons, their spin properties into applications.
As one major branch, magnetic domain wall (DW) propagation along nanowires
has attracted considerable attention in recent years due to its potential
in achieving, for instance, high-intensity information storage,
nonvolatile random access memory and DW logic circuit
\cite{Walker,Parkin,Allwood,Cros_v,Cowburn,Erskine,Wang}.

It has already been known for almost 40 years that the 1+1 (spatial and
time) dimensional Landau-Lifshitz-Gilbert (LLG) equation \cite{Landau},
which universally governs magnetization dynamics, admits a well-known
exact Walker propagating DW solution for a biaxial nanowire.
It predicts that, in the presence of an external magnetic field,
the DW subjects to a rigid-body translational motion which is
valid when the magnetic field is in a proper regime.
Despite its attractive simplicity and elegance, and the fact that
this Walker solution has played a pivotal role in our current
understanding of both current-driven and field-driven DW propagation
in magnetic nanowires \cite{Zhang,Fert,Linder,Wieser,Xiansi},
whether or not this solution is genuine, i.e., describing a
realistic physical system, is still an open question.
As one necessary touchstone, genuine solution
of a physical system must be stable against small perturbations.
By now there is no proof of the stability of the Walker solution and
the validity of it for a 1D wire is always taken as self-evident.
Any deviation in experiments or numerical simulations are assumed to
be attributed to the quasi-1D nature or other effects \cite{Fert}.
However, there are signs \cite{Wieser,Xiansi} that this solution may
be unstable. For instance, in Reference \cite{Wieser}, it is shown that under a huge hard-axis
anisotropy, a DW motion damped by spin-wave emission occurs
after the field exceeds a critical value. In addition, only stern
waves were observed therein. In Reference \cite{Xiansi}, a propagating DW
dressed with spin-waves was also captured both in the absence
and presence of the Gilbert damping, and unlike \cite{Wieser},
the spin-waves observed emit both stern and
bow waves. Moreover, apparent deviation of DW velocity and
deformation of DW profile from Walker
predicted values were also observed. Unlike the microscopic
DW profile which is sensitive to any errors incurred in
simulations, the speed of DW manifests collective behavior
of spins composing the DW; thus it is capable of reflecting
the macroscopic physics that are invulnerable to the self-averaging
microscopic perturbations when a large number
of spins are involved. Therefore, this velocity
deviation, as a more conspicuous fingerprint of the DW��s
destabilization, shall also be addressed in regards to its origin
in order for a deep understanding of DW propagation in nanowires.
On the other hand, applications of spintronics devices require
accurate description of DW motion \cite{Stohr,Gerrit1,Han,magnon}.
Thus, the stability of the Walker propagating DW solution becomes
vital in our understanding of DW propagation along a magnetic wire.
\begin{figure}
\includegraphics[width=3.4 in]{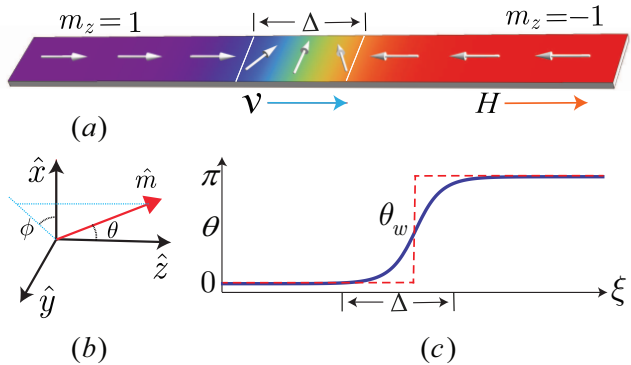}\\
\caption{(Color online) (a) Illustration of transverse head-to-head DW
of width $\Delta$ in a nanowire, with easy axis along $\hat z$ and
hard axis along $\hat x$. In the absence of external magnetic field
(upper), a static DW exists between two domains with $m_z=\pm 1$.
Under a field parallel to the easy axis, the Walker propagating DW
moves towards the energy minimum state ($m_z=-1$) at a speed $v$
while the DW profile is preserved. (b) Illustration of coordinate
system. (c) Solid line:DW profile $\theta_w$ vs. $\xi$ that varies noticeably
around the DW center. Dashed line: local perturbation to $\theta_w$.}
\label{fig1}
\end{figure}

However, unlike stability analysis of solutions of linear and nonlinear
ordinary differential equations which can be easily done by using the
linearization techniques and Lyapunov-exponent concept \cite{bhu1,xrw1},
it is hard in general for nonlinear partial differential equations like
the LLG equation. Although numerical approaches can provide clues and
hints, an analytical approach is lack for Walker solution until recently
the progress of traveling wave analysis suggests a feasible way.
In this paper, we shall present in detail the method and results of
our stability analysis of the Walker exact propagating DW solution of
a 1+1 dimensional LLG equation \cite{bhu2}. It is shown that
a propagating DW is always dressed with spin waves so
that the Walker solution is not stable against spin-wave
emission. In the low field region, only stern spin waves are
emitted while both stern and bow waves emerge under high
field. When the transverse magnetic anisotropy is larger than
a critical value and the external field is sufficiently high, the
solution is convective or absolute unstable, corresponding to
severe distortion of the propagating DW profile. This shall
lead to noticeable deviation of DW speed from the Walker
formula besides that the DW is dressed with spin waves.
The paper is organized as follows. The model and theoretical
formulation are explained in the next section.
Section III is the results and discussions, and the conclusion is
in Section IV.

\section{II. MODEL AND THEORETIC FORMULATION}

To study the stability of Walker exact propagating DW solution under an
external field, we consider the dimensionless LLG equation \cite{Wang},
\begin{equation}
\frac{\partial \vec m}{\partial t} = -\vec m\times{\vec h_{eff}}
+ \alpha \vec m \times \frac{{\partial \vec m}}{{\partial t}}.
\label{llg} \end{equation}
This LLG equation describes the dynamics of the magnetization $\vec M$
of a magnetic nanowire schematically shown in Fig. \ref{fig1}.
With the easy axis along the wire ($\hat z$ direction) and the width
and thickness being smaller than the exchange interaction length,
exchange interaction dominates the stray field energy caused by magnetic
charges on the edges; the DW structure tends to be homogeneous in the
transverse direction \cite{Porter}, i.e., behaves effectively 1D.
We are interested in the behavior of a head-to-head DW under an
external field shown in Fig. \ref{fig1}. In Eq. \eqref{llg}, $\vec m$
is the unit direction of the local magnetization $\vec M= \vec m M_s$
with saturation magnetization $M_s$ and $\alpha$ is the
phenomenological Gilbert damping constant. The effective field (in
units of $M_s$) is $\vec h_{eff}=K_\parallel m_z \hat{z}-K_\perp m_x
\hat{x}+A \partial^2\vec m/\partial z^2 +H\hat{z}$ where $K_\parallel$,
$K_\perp$, and $A$ are respectively the easy axis anisotropy coefficient,
the hard axis anisotropy coefficient, and the exchange coefficient.
$H$ is the external magnetic field parallel to $\hat{z}$.
The time unit is $(\gamma M_s)^{-1}$, where $\gamma$ is the gyromagnetic
ratio. Using polar angle $\theta$ and azimuthal angle $\varphi$ for
$\vec m$ as shown in Fig. \ref{fig1}, this LLG equation has a well
known Walker propagating DW solution \cite{Walker},
\begin{equation}
 \begin{split}
\sin 2{{\varphi_w(z,t)}} =\frac{H}{H_c}, \quad
\ln \tan \frac{1}{2}{\theta_w(z,t)}  =  \frac{z-vt}{\Delta}.
\end{split}
\label{walker} \end{equation}
Here $H_c=\alpha K_\perp /2$ is the Walker breakdown field and $\Delta=
{(K_\parallel/A + {\cos^2}{\varphi_w}K_\perp/A)^{-1/2}}$ is the DW width
which will be used as the length unit ($\Delta=1$) in the analysis below.
$v=\Delta H/\alpha$ is the Walker rigid-body DW speed that is linear
in the external field and the DW width, and inversely proportional to the
Gilbert damping constant. Solution \eqref{walker} is exact for $H<H_c$.

In the following analysis, the meaning of stability/instability
of the DW is confined to Lyapunov definition, i.e., the
DW is stable if any other solution of Eq. \eqref{llg}
starting close enough to the Walker solution will remain
close to it forever; otherwise it is unstable. We will prove
the instability of solution \eqref{walker} against spin-wave
emission by performing a spectrum analysis according to a
recent developed theory for a general travelling front,
such as a propagating head-to-head DW shown in Fig. \ref{fig1}.
To prove the instability of solution \eqref{walker},
we follow a recently developed theory
(Sandstede and Scheel \cite{Sandstede} and Fiedler and Scheel
\cite{Fiedler}) for stability of a general traveling front,
that is, a solution connecting two homogeneous states,
such as a propagating head-to-head DW shown in Fig. 1.
A {\it modus operandi} is to perturb Eq. \eqref{llg} by a small
deviation $\vec{\delta}$ from the solution \eqref{walker} via
which an evolution equation governing this deviation can be derived.
Note that if we directly perturb Eq. \eqref{llg} by $m_x^w+\delta_x$,
$m_y^w+\delta_y$ and $m_z^w+\delta_z$, with $|\delta_{x,y,z}|\ll 1$
($m_x^w$, $m_y^w$ and $m_z^w$ are components of Eq. \eqref{walker}
in the Cartesian coordinates), the three components of
$\vec{\delta}$ were not independent due to the preservation of
$\text{ }\!\!|\!\!\text{ }\vec{M}\text{ }\!\!|\!\!\text{ }$.
A convenient way to circumvent this problem is,
instead of analyzing in the Cartesian space, to work
with the polar-coordinate form of Eq. \eqref{llg} in which
the two variables $\theta $ and $\phi $ satisfy \cite{Walker}:
\begin{equation}
\begin{array}{l}
\dot \theta  - \alpha \sin \theta \dot \varphi  =
- 2{K_ \bot }\sin \theta \sin \varphi \cos \varphi  + 4A\theta '\varphi '\\
\;\;\;\;\;\;\;\;\;\;\;\;\;\;\;\;\;\;{\rm{ + }}2A\sin \theta \varphi '',\;\;\;\;\;\;\;\;\;\\
\sin \theta \dot \varphi  + \alpha \dot \theta  =
- 2{K_ \bot }\sin \theta \cos \theta {\cos ^2}\varphi  + 2{K_{//}}\sin \theta \cos \theta \;\;\\
\;\;\;\;\;\;\;\;\;\;\;\;\;\;\;\;\;\; + H\sin \theta {\rm{ + }}2A\sin \theta \cos \theta {{\varphi '}^2} - 2A\theta '',
\end{array}
\label{walkerspheric} \end{equation}
where single and double prime denote the first and the second
derivatives with respect to $z$. By assuming $\theta_w+\theta$
and $\varphi_w+\varphi$ the solution of Eq. \eqref{walkerspheric}
with $|\theta|$, $|\varphi|\ll 1$, and by keeping only the terms
of the first order in $\theta $ and $\varphi $, the linearized
equations of $\theta $ and $\varphi $ in the moving DW frame of
velocity $v$ (with the coordinate transformation $z\to \xi $ and
$t\to t$, where $\xi =z-vt$) are, in a two-component form of
$\Lambda \equiv {{(\theta ,\ \varphi )}^{T}}$
(superscript $T$ means transpose),
\begin{equation}
\frac{{d\Lambda }}{{dt}} = {L_0}\Lambda  + {L_1}\frac{{\partial \Lambda }}
{{\partial \xi }} + {L_2}\frac{{{\partial ^2}\Lambda }}{{\partial {\xi ^2}}},
\label{linearized1} \end{equation}
where $L_0$, $L_1$, and $L_2$ are $2\times 2$ matrices that depend on
$\xi $ through ${{\theta }_{w}}$.
$L_0$ has the following matrix elements: $L_{0,11}=\{\alpha K_\parallel
\cos [2G(\xi)]+K_\bot(\sqrt{1 -{\rho^2}}-1)
\cos [2G(\xi)]/2 +(H\alpha -K_\bot\rho /2)\tanh\xi \}/(1 + \alpha^2)$,
$L_{0,21}=\{ K_\parallel\cosh \xi\cos [2G(\xi)]+K_\bot(\sqrt{1-\rho^2}-1)
\cosh\xi\cos [2G(\xi)]/2+(H +K_\bot\alpha\rho /2)\sinh\xi \} /(1+\alpha^2)$,
$L_{0,12}=$sesh$\xi K_\bot(-\sqrt{1-\rho^2}+\alpha\rho\tanh\xi)/(1+\alpha^2)$,
$L_{0,22}=K_\bot (\alpha\sqrt{1 -\rho^2}+\rho\tanh \xi)/(1+\alpha^2)$.
Here $G(\xi )$ is the Gudermannian function and $\rho =H/H_c$.
$L_1$, $L_2$ can be expressed explicitly in terms of
$\xi $ as:
\[\begin{array}{l}
L_1{\rm{ = }}\left( {\begin{array}{*{20}{c}}
v&{ - \frac{{2A}}{{(1 + {\alpha ^2})\cosh \xi }}}\\
0&{v + \frac{{2A\alpha }}{{(1 + {\alpha ^2})}}}
\end{array}} \right),\;\;\;\;\;\\
L_2{\rm{ = }}\frac{1}{{(1 + {\alpha ^2})}}\left( {\begin{array}{*{20}{c}}
{A\alpha }&{ - \frac{A}{{\cosh \xi }}}\\
{A\cosh \xi }&{A\alpha }
\end{array}} \right).
\end{array}\]
Eq. \eqref{linearized1} is a linearized equation, and its general
solutions are linear combinations of basic solutions of the form,
\begin{equation}
\Lambda (\xi ,t) = {\Lambda _1}(\xi ){e^{\lambda t}},
\label{basicsolution} \end{equation}
where $\lambda $ is a proper complex number that supports
nontrivial solutions (not constant zero) for equation
\begin{equation}
(L - \lambda ){\Lambda _1}\left( \xi  \right) = 0,
\label{eigeneq} \end{equation}
where $L=L_0+L_1\partial/\partial\xi+L_2\partial^2/\partial\xi^2$.
Then all such $\lambda$ define the spectrum of $L$.
It is straightforward to verify that, due to translational
invariance of solution \eqref{walker}, $\lambda =0$
always belongs to the spectrum, with the
corresponding eigenfunction ${\Lambda _1} =
{({{\partial {\theta _w}}}/{{\partial \xi }},
{{\partial {\varphi _w}}}/{{\partial \xi }})^T}.$
If none of $\lambda$ in the spectrum has positive real part,
the spectrum is said to be stable; otherwise it is unstable.
For a stable spectrum, any moderate deviations from the Walker
solution must either decay exponentially with time
[$\operatorname{Re}(\lambda )<0$] or undergo periodic motion by
retaining its amplitude [$\operatorname{Re}(\lambda )$$=0$].
When the spectrum encroaches the left half plane, exponentially
growing modes ($\operatorname{Re}(\lambda )$$>0$) exist.
We shall use so-called essential and absolute spectra of $L(\xi)$
to decide the stabilities/instabilities of domains and DW profile.

\section{III. RESULTS AND DISCUSSIONS}

\subsection{A. ESSENTIAL INSTABILITY}

In order to compute the spectrum of $L(\xi)$, it is convenient to
rewrite Eq. \eqref{eigeneq} in the first order differential form by
using $\Omega\equiv (\theta, \;\varphi,\;\partial \theta /\partial
\xi ,\;\partial \varphi /\partial \xi )^T$,
\begin{equation}
\frac{d}{{d\xi }}\Omega = \Gamma (\lambda ,\xi )\Omega,
\label{4ode} \end{equation}
where
\begin{equation}
\Gamma (\lambda) = \left( {\begin{array}{*{20}{c}}0&I\\
L_2^{-1}(\lambda-L_0)&-L_2^{-1}L_1 \end{array}} \right).
\label{linearize3}
\end{equation}
$I$ is the $2\times 2$ identity matrix.
All $\lambda$ that supports nontrivial solutions to Eq.
\eqref{4ode} form its spectrum. Eq. \eqref{eigeneq} and
Eq. \eqref{4ode} have the same spectrum because they are equivalent.
We shall focus hereafter on the spectrum of Eq. \eqref{4ode}.
To do so, we need to obtain the conditions under which Eq. \eqref{4ode}
has nontrivial solutions. Let us first divide $\xi$ axis into four
regions: $\xi\le -l$, $-l\le \xi\le 0$, $0\le \xi\le l$ and $\xi\ge l$
with $l\gg \Delta$.
Notice that $\Gamma$ depends on $\xi$ only through $\theta_w$ that
varies with $\xi$ only within the DW, Eq. \eqref{4ode} is essentially,
\begin{equation}
\frac{d}{d\xi }\Omega = \Gamma^- (\lambda)\Omega.
\label{4odelimit} \end{equation}
in region $-\infty<\xi\le -l$ and
\begin{equation}
\frac{d}{d\xi }\Omega = \Gamma^+ (\lambda)\Omega.
\label{4odelimit2} \end{equation}
in region $l\le\xi<\infty$. The two asymptotic matrices $\Gamma^\pm$ are,
\begin{equation}
\Gamma^\pm \left(\lambda\right) =\lim_{\xi\to\pm\infty}\Gamma(\lambda,\xi).
\label{limitmatrix} \end{equation}
$\Gamma^\pm$ can be directly obtained from Eq. \eqref{linearize3} by
replacing $\theta_w(\xi)$ with $\pi$ for $+$ and with $0$ for $-$.
In region $-\infty<\xi\le -l$ ($l\le\xi<\infty$) and for each given
$\lambda$, $\Gamma^\pm (\lambda)$ has 4 eigenvalue and eigenvector pairs,
$(\kappa^\pm_i, \mu^\pm_i)$ $i=1,\ldots,4$, and Eq. \eqref{4odelimit}
(\eqref{4odelimit2}) has solution of form $\mu_i^- e^{\kappa_i^- \xi}$
($\mu_i^+ e^{\kappa_i^+ \xi}$). $\lambda$ can then be denoted by
$(n^\pm_+,n^\pm_-)$ for $n^\pm_+$ ($n^\pm_-$) being the number of
$\kappa^{\pm}(\lambda)$ with positive (negative) real parts.
Obviously, we have $n^+_++n^+_-=n^-_++n^-_-=4$ except on
the so-called Fredholm borders explained below in detail.
$\kappa^\pm$ can then be ordered descending by their real parts as
${\mathop{\rm Re}\nolimits} (\kappa _1^ \pm ) \ge ... \ge {\mathop{\rm Re}
\nolimits} (\kappa _{n_ + ^ \pm }^ \pm ) > 0 > {\mathop{\rm Re}\nolimits}
(\kappa _{n_ + ^ \pm  + 1}^ \pm ) \ge ... \ge {\mathop{\rm Re}\nolimits}
(\kappa _4^ \pm )$.
Each solution $\mu_i^-e^{\kappa_i^-\xi}$ ($\mu_i^+e^{\kappa_i^+\xi}$)
in region $-\infty<\xi\le -l$ ($l\le\xi<\infty$) can be continued
into region $-l\le\xi\le 0$ ($0\le \xi\le l$) as $\Omega_i^(\xi)$
($\Omega_i^+(\xi)$).
Suppose we are interested in a nontrivial bounded solution $\Omega$ of
Eq. \eqref{4ode}, i.e., $\Omega(\pm \infty)=0$, then $\Omega$ must be
the linear superposition of those $\Omega_i^-(\xi)$ [$\Omega_i^+(\xi)$]
in $-l\le\xi\le 0$ ($0\le \xi\le l$) whose corresponding eigenvalues
$\kappa_i^-$ ($\kappa_i^+$) have positive (negative) real parts.
Note that the number of $\kappa_i^-$ ($\kappa_i^+$) with
Re$(\kappa_i^-)>0$ (Re$(\kappa_i^+)<0$) is $n^-_+$ ($n^+_-$),
whether or not such $\Omega$ exists is equivalent to whether
or not we can find nontrivial solution $(a_i,b_j)$ satisfying
\begin{equation}
\sum\nolimits_{i=n_+^++1}^{4} a_i\Omega_i^+(0) =
\sum\nolimits_{j=1}^{n_+^-} b_j\Omega_i^-(0).
\label{continuity} \end{equation}
This is the condition of the continuation of $\Omega$ at $\xi=0$.
The spectrum of Eq. \eqref{4ode} is the set of all $\lambda $ such
that Eq. \eqref{continuity} has at least one nonzero solution of
$(a_i,b_j)$ for $i=n_+^-+1\ldots 4$ and $j=1\ldots n_+^-$.
Obviously, there are $n_-^++n_+^-$ variables and 4 equations.
The existence of such a solution is then $n_-^++n_+^->4$.
The explict solutions of $(a_i,b_j)$ require the knowledge of
$\Omega_i^\pm(0)$ that is normally not known analytically because
of the complicate $\xi$-dependence of $\Gamma (\lambda,\xi)$.
Numerical method such as the shooting algorithm used in the
Schrodinger equation may be use here by numerically integrating
Eq. \eqref{4ode} starting from $\xi=\pm l$ (where all linear independent
solutions of Eqs. \eqref{4odelimit} and \eqref{4odelimit2} are known) and
ending at $\xi=0$. Correct set of $(a_i,b_j)$ shall make the shooting
of $\Omega$ from $\xi=\pm l$ end with the same value at $\xi=0$.
The shooting algorithm, proved to be efficient for the Schrodinger equation
whose spectrum is on the real line, may become excessively arduous
for the LLG Eq. where the spectrum extends to the whole complex plane.
As we shall see, this formidable task can be partly dodged as far as only
the essential instability is considered which is pertinent to spin wave
emissions.

Similar to the energy spectrum of a quantum system, the spectrum
$\lambda$ of Eq. \eqref{4ode} can be discrete and continuum.
The continuum $\lambda$ is also called the essential spectrum.
The essential spectrum is not sensitive to the so called
relatively compact perturbations to Eq. \eqref{4ode}.
Here a relatively compact perturbation can be understood, in some
senses, as a local perturbation to a Schrodinger equation
\[[ - \frac{{{d^2}}}{{d{x^2}}} + V(x)]\psi  = E\psi .\]
This continuous spectrum will not be changed by a $V(x)$ of finite
potential range, such as a potential well or barrier, although
wave functions are altered and point spectrum may be introduced.
According to \cite{nbook,Henry,Fiedler,Sandstede}, a similar local
perturbation to Eq. \eqref{4ode} (or in general to any linearized
equation of a system around a front solution) preserve the essential
spectrum so that we can replace $\theta_w$ (Fig. \ref{fig1} (c)) by
$\pi H(\xi)$, where $H(\xi)$ is the Heaviside step function.
The new equation with the same essential spectrum as that of
Eq. \eqref{4ode} is
\begin{equation}
\frac{d}{{d\xi }}\Omega = \Gamma^\infty \Omega,
\label{asymptotic}
\end{equation}
where
\begin{equation}
{\Gamma ^\infty } \equiv \left\{ \begin{array}{l}
{\Gamma ^ +(\lambda) },\;\;\;\;\;\;\xi  \ge 0\\
{\Gamma ^ -(\lambda) },\;\;\;\;\;\;\xi  < 0
\end{array} \right.
\label{linearize4}
\end{equation}
$\Gamma^\pm(\lambda)$ have already been defined in Eq. \eqref{limitmatrix}.
Since $\Gamma^\infty$ is constant in each region of $(-\infty ,0)$ and
$(0,\infty )$, the corresponding $\Omega_i^\pm$ in Eq. \eqref{continuity}
are just eigenvectors $\mu_i^\pm$ of $\Gamma^\pm$.
Therefore Eq. \eqref{continuity} becomes
\begin{equation}
\sum\nolimits_{i=n_+^++1}^4a_i\mu_i^+ =
\sum\nolimits_{j=1}^{n_+^-}b_j\mu_j^- .
\label{eqs2}
\end{equation}
Eq. \eqref{eqs2} have nonzero solution if the number of
variables $a_i$ and $b_j$, $n_-^+ -n_+^-$, is greater than 4.
Thus, Eq. \eqref{asymptotic} has nontrivial solution bounded at
$\xi=\pm \infty$ for all $\lambda$ whose $n_-^+ +n_+^->4$.
If one allow other types of solutions at $\xi=\pm \infty$, then
the general condition is $n_-^+ +n_+^-\neq 4$. Indeed according to the
theory of Refs. \cite{Sandstede,Palmer,Henry,Fiedler,nbook},
the essential spectrum of $L$ (also of Eq. \eqref{4ode} and/or Eq.
\eqref{asymptotic}) is the union of all closed sets of $\lambda$
(boundaries included) whose indices $n_-^+$ and $n_+^-$
satisfy $n^+_-+n^-_+ \neq 4$.
The boundaries of each region, known as the Fredholm borders,
must be those lines crossing which either $n^+_-$ or $n^-_+$
changes its value by 1.
Then along each Fredholm border, either $\Gamma^+$ or $\Gamma^-$
must possess pure imaginary eigenvalue (not a hyperbolic matrix);
thus these lines can be determined by
$\det[\Gamma^{\pm}(\lambda)+ik]=0$  with
$k\in(-\infty,\infty)$ \cite{Sandstede,Palmer,Fiedler}.
Each of the two equations has two branches of allowed $\lambda$
denoted as $\lambda_{1,2}^{\pm}(k)$.
Note that Eqs. \eqref{4odelimit} and \eqref{4odelimit2} admit pure
plane wave solution $\Omega_0 e^{ik \xi}$ when $\lambda$ is on
$\lambda_{1,2}^+(k)$ ($\lambda_{1,2}^-(k)$). Therefore an encroachment
of these borders to the right half plane implies spin wave emission.
We refer to the type of instability characterized by the presence of
essential spectrum on the right half plane as the essential instability.

\begin{figure}
  \includegraphics[width=3.4 in]{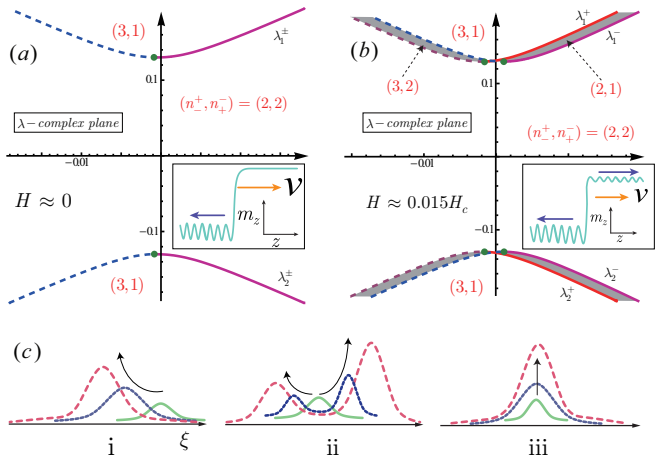}\\
\caption{(Color online) Left are the essential spectrum (shadowed regions)
for $H=0$ (a) and $H \approx 0.015 H_c$ (b). The Fredholm borders are
$\lambda_{1,2}^{\pm}(k)$. Solid border lines correspond to spin waves
with negative group velocities  while the dashed border lines
are for the spin waves with positive group velocities.
Propagating DW wall emit stern waves in low fields [right of (a)], and
stern and bow waves in higher field ($0.015H_c<H<H_c$) [right of (b)].
The green dots are zero group velocity modes. $K_\perp=0.4$ is used.
(c) Illustration of transient instability (i) in which the unstable mode
travels to only one direction; convective instability (ii) in which
the unstable mode  travels to both directions; and the absolute
instability (iii) in which the unstable mode does not travel (stationary
in the moving frame). }
\label{fig2}
\end{figure}

In order to understand numerical results in Ref. \cite{Xiansi}, parameters
of yttrium iron garnet (YIG) \cite{magnon} are assumed in our analysis with
$A= 3.84 \times 10^{-12}  J/m$, $K_{\parallel}=2 \times 10^3 J/m^3$,
$\gamma=35.1 kHz/(A/m)$, and $M_s=1.94 \times 10^5 A/m$.
$\alpha=0.001$ is used and $K_{\perp}$ is a varying parameter.
Fig. \ref{fig2} plots the essential spectrum for $K_\perp=0.4$ (in units of
$\mu_0M_s^2$ that is about 10 times larger than $K_{\parallel}$).
The qualitative results are very similar to the early results \cite{bhu2}:
In the absence of an external field, the two branches of the spectrum of
$\Gamma^\pm$ are the same, $\lambda_{1,2}^{+}(k)=\lambda_{1,2}^{-}(k)$,
shown in Fig. \ref{fig2}(a). Since the spectrum encroaches the right half
plane, unstable plane waves shall exist and spin wave emission are expected.
Similar conclusion was also obtained in early study \cite{Bouzidi}, but for
$H>H_c$. Solid lines are for negative group velocity [determined by
Im$({\partial \lambda}/{\partial k})$], thus these are stern modes.
The dashed lines indicate positive group velocity, corresponding to
bow modes. The green dots are zero group velocity points.
According to Fig. \ref{fig2}(a), all unstable modes have negative group
velocities so that DW can only emit stern waves in the low fields.
As the external field increases, $\lambda_{1,2}^{+}(k)$ and $\lambda_{1,2}
^{-}(k)$ will separate, and the area of essential spectrum in $\lambda
$-plane becomes bigger and bigger (shadowed regimes in Fig. \ref{fig2}(b).
The green dots also moves toward Im$(\lambda)$-axis and cross it at
$H \approx 0.015H_c$ [Fig. \ref{fig2}(b)].
Upon further increase of $H$, the unstable modes have both positive and
negative group velocities although the most of them have the negative ones.
One shall have propagating DW to emit both stern and bow waves.
The stern waves should be stronger than the bow waves as schematically
shown in the right figure of Fig. \ref{fig2}(b). This is exactly what were
observed in numerical simulations for stern wave emission in low field
\cite{Wieser} and stern-and-bow wave emission in high field \cite{Xiansi}.
In a realistic wire with damping, emitted spin waves will be dissipated
after a short distance, and are hard to be observed in experiments.

\subsection{B. TRANSIENT/CONVECTIVE/ABSOLUTE INSTABILITY}

The essential spectrum decides the instability of domains.
DW propagation will generate spin waves in domains when the
essential spectrum encroaches the right half of the $\lambda$ plane.
The fact that the essential spectrum is not affected by the variation
of DW profile means that the essential spectrum cannot determine
the instability of DW profile that is important for many quantities
such as the DW velocity. Interestingly, DW instability is determined
by the so-called absolute spectrum explained below.
It can be classified into three categories. Absolute instability [AI,
Fig. \ref{fig2}(c)iii] occurs when at each fixed point on the
$\xi$-axis, the disturbance grows exponentially with time.
It is associated with the emergence of nontraveling unstable modes in the
absolute frame (the moving frame that we adopted); thus coins its name.
This point-wise growth feature of AI is in sharp contrast with the other
two types of instability, which albeit grows in the total norm,
decays locally at each fixed point on the $\xi$-axis.
They happens when all unstable modes are transported to infinities at
fast enough velocities. It is called a transient instability (TI,
Fig. \ref{fig2}(c)i) if the disturbance generated locally in the DW
region transports to infinity in one direction (either towards
$\infty$ or $-\infty$), while it is called convective (CI, Fig.
\ref{fig2}(c)ii) if it can transport in both directions.
Intuitively, transient instability shall have the least influence on the
DW property since once generated, it will leave the DW region quickly
and will not interact with the DW hereafter. Convective instability is
stronger than the transient one since although transported outside the
DW region, it could influence the DW through second order effect in
which new bidirectional unstable modes excited by the convecting wave
packets can collide and interact with the DW again. Absolute instability
is the most severe one in the sense that once a nontraveling disturbance
is generated, it can stay within and keep interacting with the DW,
leading to dramatic modification on the DW profile. For this reason,
physical quantities depending on DW profile, such as the DW velocity,
are expected to be strongly affected.

The three types of transportation behavior, either unidirectional,
bidirectional or non-travelling, are determined by the so-called
absolute spectrum and the branching points
\cite{Sandstede,Palmer,Henry,Fiedler,nbook,Sandstede2,Chomaz,Brevdo}.
To introduce the absolute spectrum and the branching points, we recall
that, for each $\lambda$ in the complex plane, there are four $\kappa_
i^\pm$ ($i=1,2,3,4$) for $\Gamma ^\pm$, ordered by their real parts as
$\hbox{Re}(\kappa_1^\pm)\ge \hbox{Re}(\kappa_2^\pm \ge \hbox{Re}
(\kappa_3^\pm)\ge \hbox{Re}(\kappa_4^\pm)$.
Then $\lambda$ is said to belong to the absolute spectrum ($\lambda_{abs}$) if and only if $\hbox{Re}[\kappa_2^+(\lambda)]=\hbox{Re}[\kappa_3^+(\lambda)]$
or $\hbox{Re}[\kappa_2^-(\lambda)]=\hbox{Re}[\kappa_3^-(\lambda)]$.
The branching points are special points in the absolute spectrum,
denoted as $\lambda_{sd}$, satisfying $\kappa_2^\pm (\lambda_{sd})
=\kappa_3^\pm (\lambda_{sd})$. To have a better feeling about the
absolute spectrum and differences in unidirectional/bidirectional
transportation and nontraveling modes of a wave $\Lambda(z,t)$,
we introduce the concept of pointwise decay and growth.
A wave $\Lambda(z,t)$ is said to be pointwise decay iff $\lim_{t\to\infty}\Lambda (z_0)=0$ for any fixed $z_0$.
The opposite ($\infty$ instead of 0) is said to be pointwise growth.
Let us first consider a wavelet that may exemplify a transient
disturbance transporting to the right along the z axis:
\begin{equation}
\Lambda= e^{\lambda t}\hbox{sech} (z-vt).
\label{wavelet}
\end{equation}
This is an unstable mode if $\hbox{Re}(\lambda)>0$.
At each fixed point $z_0$, $\lim_{t\to\infty}\Lambda (z_0)=0$ ($\infty$)
if $v>\hbox{Re}(\lambda)$ ($v<\hbox{Re}(\lambda)$).
In another word, an unstable disturbance moving fast enough can lead to
pointwise decay (vanish in a long time at each fixed point) although
its norm $\| \Lambda\|=\int_{-\infty}^\infty\left|\Lambda\right|^2 dz
=\pi e^{\lambda t}$ increases exponentially with time.
Interestedly, \eqref{wavelet} can be brought to be stable when
$v>\hbox{Re} (\lambda)$ if an exponential weight $e^{\eta z}$
is used
\begin{equation}
\| \Lambda\|_\eta=\int_{-\infty}^\infty\left|
e^{\eta z}\Lambda \right| ^2 dz=e^{\lambda't}C_0,
\label{expweight}
\end{equation}
where
\[\lambda'=2(\lambda  + \eta v)\]
and
\[C_0=\int_{-\infty}^\infty e^{2\eta z'}{\mathop{\rm sech}\nolimits
}^2(z')dz' .\]
Note that the integral $C_0$ is finite whenever $|\eta|<1$.
Therefore any $-1<\eta<-[\hbox{Re}(\lambda)/v]$ makes
$\hbox{Re}(\lambda')<0$ such that the new norm \eqref{expweight}
decay exponentially with time.
However, if $v<\hbox{Re}(\lambda)$, either \eqref{expweight}
diverges with time or $C_0$ is infinity for any $\eta$.
In another word, the mode becomes stable under a proper exponential
weight $\|.\|_\eta$ for $v>\hbox{Re}(\lambda)$, and a transient
disturbance traveling towards $-\infty$ fast enough can be stabilized
by a positive $\eta$ since then the multiplier $e^{\eta z}$ balance
the growing modes at $-\infty$.
\begin{figure}
  \includegraphics[width=3.4 in]{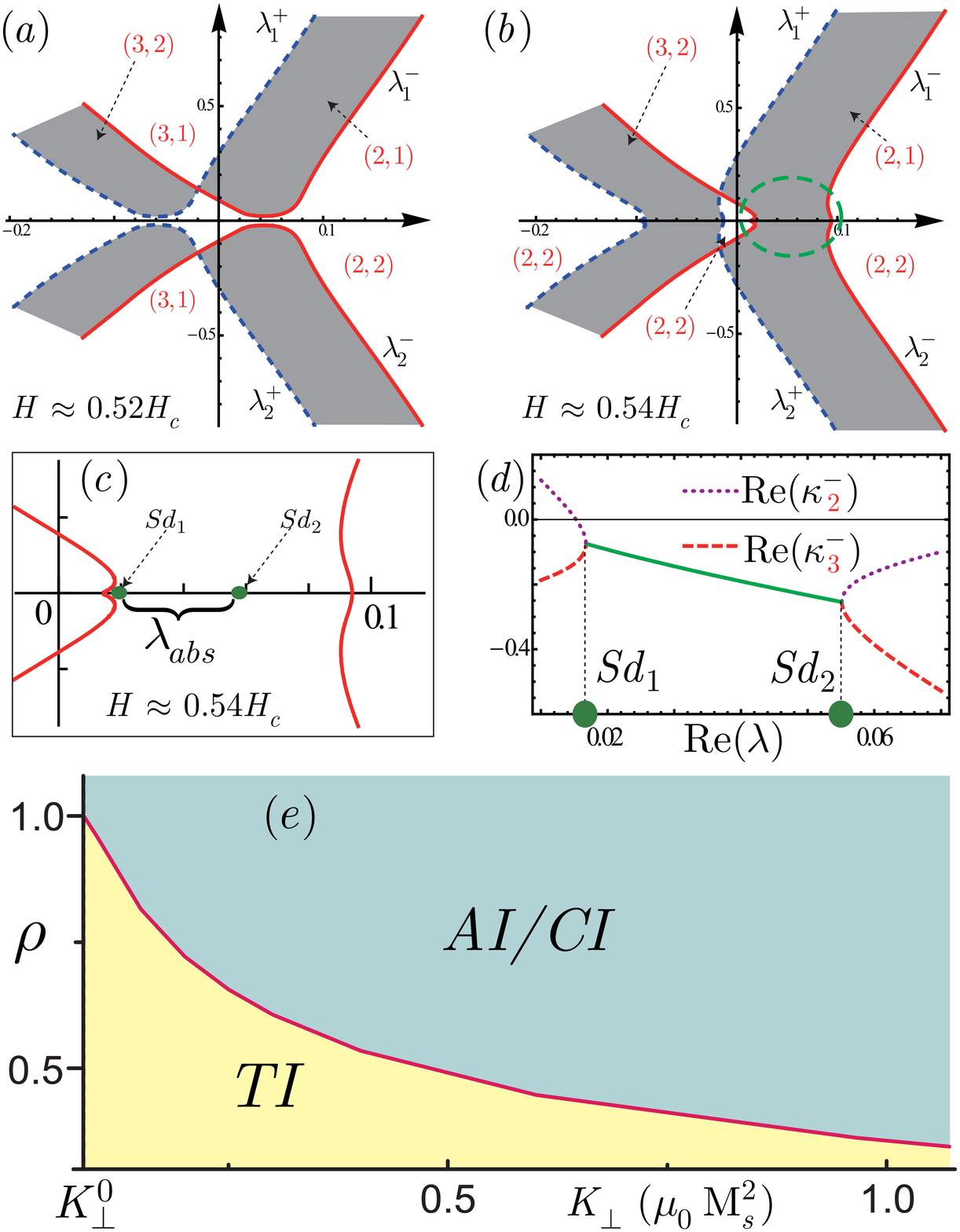}\\
\caption{(Color online)
Essential spectra and $(n^+_-,n^-_+)$ in regions divided by $\lambda
^\pm_{1,2}$ for $K_\perp=0.4$, $\rho=0.52$ (a) and $0.54$ (b).
No absolute spectrum presents before $\lambda^\pm_{1,2}$ tangents
at the real axis while in (b), unstable absolute spectrum presents
in the region enclosed by the solid circle.
(c) Enlarged description of the region enclosed by the solid circle
in (b). The absolute spectrum is between the two branching points
$Sd_1$ and $Sd_2$ (green dots). (d) Plot of $\hbox{Re}(\kappa^-_2)$
and $\hbox{Re} (\kappa^-_3)$ vs. $\lambda$ between $Sd_1$ and $Sd_2$.
At $Sd_{1,2}$, $\kappa^-_2=\kappa^-_3$. (e) Phase diagram of
transient (TI) and absolute/convective (AI/CI) instabilities.
The boundary is the bifurcation line between TI/CI-and-CI
instabilities in $K_\perp$ and $\rho=H/H_c$ plane.
The bifurcation line is only plotted for $K_\perp \geq K_\perp^0 $
here $K_\perp^0\approx 0.085$ at which $H_2=H_c$ ($\rho=1$).
Noted that our analysis is valid for fields below the Walker
breakdown value.}
\label{fig3}
\end{figure}

In general, the exponentially weighted norm denoted by
$\left\| \cdot \right\|_\eta$ for a real number $\eta$ is defined as
\begin{equation}
\left\|\Lambda\right\|_\eta\equiv\int_{-\infty}^\infty\left|
e^{\eta\xi}\Lambda \right|^2 d\xi .
\label{weight}
\end{equation}
The transient and convective instabilities behave very differently under
the norm. For a given $\lambda$, its eigenmode is transient unstable if
it has an exponentially growing factor that travels towards $-\infty$
(or $\infty$). Under an exponentially weighted norm with a proper choice
of $\eta>0$ ($\eta<0$) for mode traveling to $-\infty$ ($\infty$), the
growth at $-\infty$ ($\infty$) can be absorbed by the multiply $e^{\eta\xi}$.
Therefore the essential spectrum calculated under the exponential norm
With $\eta>0$ ($\eta<0$) can be transferred to the left half of the
$\lambda$ plane for the unidirectional modes traveling towards
$-\infty$ ($\infty$). Mathematically, this corresponds to a proper
choice of the origin of the $\lambda$ plane in some sense.
Thus, with the proper definition of the norm by choosing a large enough
$|\eta(\lambda)|$, all unstable unidirectional eigenmodes of eigenvalues
$\lambda$ (essential spectrum in the right half of the $\lambda$ plane)
are removable because all such $\lambda$ can be transferred to the left
half of the $\lambda$ plane. This treatment fail to the modes traveling
to both directions of $\xi=\pm\infty$ (bidirectional eigenmodes).
They are not removable since an exponential weight can only suppress
the growth in one direction and blow up in the other direction.
The ability/inability of using an exponential weight \eqref{weight} to
stabilize/destabilize transient/convective modes leads to the following
properties: TI occurs if all unstable essential spectrum can be move to
the left half $\lambda$ plane under a proper exponentially weighted norm
while it is CI or AI if part of the unstable essential spectrum cannot
be stabilized by the norm. A naturally raised question then is which
part of the essential spectrum cannot be removed by this weight.

The answer is quite simple: The absolute spectrum cannot be moved
around in the $\lambda$ plane by introducing an exponential weight.
It must locate to the left of the rightmost Fredholm border.
If it encroaches the right half of the $\lambda$ plane, then the
essential spectrum cannot be stabilize no matter how one chose the
exponential weight $\eta$. To see why this is so, it is noticed that we
need to introduce a weight $\eta^-(\lambda)$ [$\eta^+(\lambda)$] in order
to move the Fredholm border determined by Eq. \eqref{4odelimit} [\eqref{4odelimit2}]. Thus, by using the exponential weight of
$\eta^\pm$, it is equivalent to shift the eigenvalues of $\Gamma^\pm$ \cite{Sandstede,Palmer,Henry,Fiedler,
nbook,Sandstede2,Chomaz,Brevdo} by
\begin{equation}
\kappa_i^\pm\to\tilde\kappa_i^\pm\equiv\kappa_i^\pm -{\eta^\pm},
\label{inducemap}
\end{equation}
and accordingly the indices of the $\lambda$ are transformed as
\begin{equation}
(n_-^+,n_+^-)\to (\tilde n_-^+,\tilde n_+^-),
\label{inducemap2}
\end{equation}
Now suppose $\lambda$ with $\hbox{Re}(\lambda)>0$ belongs to the
Essential spectrum but not on the Fredholm border, which means
$n_-^+ +n_+^-\neq 4$. For the LLG equation and independent of the norm
we use, $\lambda$ in the right hand side of the rightmost Fredholm
border has the indices of $n_-^+= n_+^-=2$, then all possible
combinations of $(n_-^ +,n_+^-)$ in the regions right after passing
through the rightmost Fredholm border can only be one of the four cases:
$(1,2)$, $(2,1)$, $(2,3)$, $(3,2)$. Consider for instance
$(n_-^+, n_+^-)=(1,2)$, then obviously $\hbox{Re}(\kappa^-_2)>0>\hbox{Re}
(\kappa^-_3)$ and $\hbox{Re}(\kappa^+_2),\ \hbox{Re}(\kappa^+_3)>0$.
If we also have $\hbox{Re}(\kappa^+_2) \neq \hbox{Re}(\kappa^+_3)$,
i.e. $\lambda \notin \lambda_{abs}$,
we could always find the aforementioned proper weight as, for instance:
\begin{equation}
{\eta ^{\pm}} = \left\{ \begin{array}{l}
{\eta ^{\rm{ + }}}{\rm{ = }}\frac{{{\mathop{\rm Re}\nolimits} (\kappa _2^ + ){\rm{ + }}{\mathop{\rm Re}\nolimits} (\kappa _3^ + )}}{2},\\
{\eta ^ - } = 0,
\end{array} \right.
\label{inducemap2}
\end{equation}
which means that, for essential spectrum calculated under this norm,
$\lambda$ is well to the right of the essential spectrum.
We can thus remove all unstable $\lambda$ (i.e., $\hbox{Re}(\lambda)>0$)
in this way if there is no $\lambda_{abs}$ on the right half of the
$\lambda$ plane. However, if $\lambda$ belongs to $\lambda_{abs}$ such
that $\hbox{Re}(\kappa^+_2)=\hbox{Re}(\kappa^+_3)$, it is easy to verify
that no such pair of $\eta^\pm$ exist. This absolute spectrum is exactly
the set of $\lambda$ which could not be stabilized by the aforementioned
proper weights $\eta^\pm$. Therefore we conclude that the absence of
$\lambda_{abs}$ in the right half of the $\lambda$ plane indicate
Transient instability in which all eigenmodes are unidirectional while
the presence of the absolute spectrum means emergence of bidirectional
eigenmodes.

Finally, the presence of unstable non-traveling modes is
associated with the branching points's presence on the right half plane.
It is straightforward to verify that the eigenmodes associated
with these branching points have zero group velocity, as follows.
Denote the secular polynomial of $\Gamma^\pm (\lambda)$ as:
$F(\lambda ,\kappa ) \equiv \det [\Gamma^\pm (\lambda ) - \kappa I].$
Then for $\lambda_{sd}$ satisfying $\kappa _2^ \pm ({\lambda _{sd}})
= \kappa _3^ \pm ({\lambda _{sd}})$, it must hold that:
\begin{equation}
F({\lambda _{sd}},\kappa ) = {(\kappa  - \bar \kappa )^2}(\kappa  - {\kappa _1})(\kappa  - {\kappa _4}),
\label{nontraveling}
\end{equation}
where $\bar \kappa  \equiv {\kappa _2}{\rm{ = }}{\kappa _3}$.
Then the group velocity $v$ of modes associated with $\lambda_{sd}$ is
\begin{equation}
v={\mathop{\rm Im}\nolimits} \left( {\frac{\partial \lambda }{\partial\kappa }}\right)\left| _{\kappa {\rm{ = }}\bar \kappa ,\\ \lambda  =
\lambda_{sd}} \right. = {\mathop{\rm Im}\nolimits} \left( -\frac{\partial_k F}{\partial_\lambda F} \right)\left| _{\kappa
{\rm{ = }}\bar \kappa ,\lambda  = \lambda_{sd}} \right. = 0.
\label{nontraveling2}
\end{equation}
Thus, branching points $\lambda_{sd}$ are non-travelling eigenmode
\cite{Sandstede2,Rademacher}.
For $K_\perp=0.4$, the absolute spectrum in the right half of the
$\lambda$ plane is generated by $\Gamma^-$.
Fig. \ref{fig3}(a) shows two branches $\lambda_{1,2}^-$.
They are well separated by the real axis for $\rho=0.52$ and
no absolute spectrum could be found in the right half plane.
As the field increases, the two branches get closer with each other
and at an onset field $H_2$, depending on  $K_\perp$, two branches
tangent at the real axis and then separate again in horizontal
direction as shown in Fig. \ref{fig3}(b) for $\rho=0.54$.
At this moment, unstable absolute spectrum begins to emerge
on the real axis (enclosed by the dashed circle).
Fig. \ref{fig3}(c) is the enlarged vision showing the absolute spectrum
(the segment between two branching points $Sd_{1,2}$ (green solid dots)).
The dependence of $\hbox{Re}(\kappa_2^-)$ or $\hbox{Re} (\kappa_3^-)$
on $\hbox{Re}(\lambda)$ between these two points is shown in
Fig. \ref{fig3}(d).

According to Refs. \cite{Sandstede2,Chomaz,Brevdo}, wavepackets
would be emitted if the essential spectrum encroaches the right
half $\lambda$-plane. There are three types of instability
\cite{Sandstede,Palmer,Fiedler,Henry,Sandstede2,Chomaz,Brevdo}.
The instability is called transient (TI) if the essential
spectrum encroaches the right half plane and absolute spectrum are
either in the left half plane or does not exist. The propagating DW
emits stern waves shown in Fig. \ref{fig2}(c)i. The instability is
called convective if both essential and absolute spectrum encroaches
the right half $\lambda$-plane. In this case, the emitted waves
can propagate in both direction as shown by Fig. \ref{fig2}(c)ii.
For an convective instability, if any branching point is also in
the right half $\lambda$-plane, the instability is called absolute.
An absolute instability can then emit non-traveling (zero group
velocity) waves as illustrated in Fig. \ref{fig2}(c)iii.
For LLG equation, since the absolute spectrum is the segment
connecting two branching points $Sd_{1}$ and $Sd_{2}$ [Fig. \ref{fig3}
(c), (d)], the absolute instability (AI) and convective instability (CI)
co-exist. It is known that transient instability is very weak that
can be removed under proper mathematical treatment
\cite{Sandstede,Humpherys}. Thus, we should not expect to have great
physical consequences.
On the other hand, the absolute instability move with the DW, and
cause the change of DW profile \cite{Pego,Sandstede2,Humpherys}.
It is known \cite{Wang} that field-induced DW propagating speed is
proportional to the energy damping rate that is sensitive to DW profile.
Therefore absolute instability, which deform propagating DW profile,
shall substantially alter DW speed. This may explain why the
field-induced DW speed start to deviate from the Walker result
only when the field is large enough to emit both stern and bow
waves in simulations \cite{Xiansi}.

Fig. \ref{fig3}(e) is the calculated phase diagram in
$K_\perp$ and $\rho=H/H_c$ plane. A transition from transient
instability (denoted as TI in the figure) to absolute/convective
instability (AI/CI) occur at a critical field $H_2$ as lng as
$K_\perp>K_\perp^0 \approx 0.085$ at which $H_2=H_c$.
It means no absolute/convective instability exist for
$K_\perp<K_\perp^0$, and one shall not see noticeable change in
famous Walker propagation speed mentioned early. This may
explain why many previous numerical simulations on permalloy,
which have small transverse magnetic anisotropy, are consistent
with Walker formula.
A snapshot of the convecting wavepackets could be identified in
Fig. 2 in Reference \cite{Xiansi} where wavepackets can be seen
in the vicinity of the traveling DW and travel to
both directions.

It should be noticed that the effects of point spectrum have not
been analyzed. In principle, it can also affect the stability of
the Walker solution, and should be a very interesting subject too.
Unfortunately, there are not many theorems on the point spectrum
yet. Thus, one can only rely on a numerical method to find a
point spectrum of operator $L$ and to find out whether it can
also induce any instability on a propagating DW.

\section{IV. CONCLUSIONS}

In conclusion, we present a powerful recipe for analyzing the
stability of a front of partial differential equation.
For the Walker propagating DW solution of the LLF equation in
1+1 dimension, it is found that DW will always emit stern waves
in a low field, and both stern and bow waves in a higher field.
Thus the exact Walker solution of LLG equation is not stable.
The true propagating DW is always dressed with spin waves.
In a real experiment. the emitted spin waves shall be damped away
during their propagation, and make them hard to be detected in
realistic wires.
For a realistic wire with its transverse magnetic anisotropy
larger than a critical value and when the applied external field
is larger than certain value, a propagating DW may undergo
simultaneous convective and absolute instabilities.
As a consequence, the propagating DW will not only emit
both spin waves and spin wavepackets, but also change
significantly its profile.
Thus, the corresponding Walker DW propagating speed will
deviate from its predicted value, agreeing very well with
recent simulations.

%-------------------------------------------------------------
\section{ACKNOWLEDGEMENTS}
This work is supported by Hong Kong RGC Grants (604109 and 605413).

\end{document}